\documentclass[twocolumn,aps,showpacs,floatfix]{revtex4}
 \usepackage{graphicx}
 \begin{document}
\setlength{\unitlength}{1mm}
\bibliographystyle{unsrt} 
\title{ Geometric  Phases generated by the non-trivial spatial  topology \\of
static vector fields linearly coupled to a neutral spin-endowed particle.\\Application to $^{171}$Yb atoms trapped in a 2D optical lattice}
 \author{ Marie-Anne Bouchiat}
\affiliation{Laboratoire Kastler-Brossel, CNRS, UPMC, \'Ecole Normale Sup\'erieure,   
24, rue Lhomond, 75005 Paris France,}
\author{ Claude Bouchiat}
\affiliation{Laboratoire de Physique Th\'eorique de l'\'Ecole Normale Sup\'erieure, CNRS, UPMC,  
24, rue Lhomond, 75005 Paris France.}
\date{June 14, 2012}
 \newcommand \be {\begin{equation}}
\newcommand \ee {\end{equation}}
 \newcommand \bea {\begin{eqnarray}}
\newcommand \eea {\end{eqnarray}}
\newcommand \nn \nonumber
\def \(({\left(}
\def \)){\right)}
 \def \va{{\mathbf{a}}}
 \def \vb{{\mathbf{b}}}
\def \vs{{\mathbf{s}}}
 \def  \vS{{\mathbf{S}}}    
 \def \vI{{\mathbf{I}}}
 \def \vJ{{\mathbf{J}}}
 \def \vL{{\mathbf{L}}}
 \def \vr{{\mathbf{r}}}
 \def \vF{{\mathbf{F}}}
 \def \vr{{\mathbf{r}}}
\def \vp{{\mathbf{p}}}
\def \vE{{\mathbf{E}}}
 \def \vbE{{\mathbb{E}}}
\def \vX{{\mathbf{X  }}} 
\def \vB{{\mathbf{B}}}
\def \ve{{\mathbf{e}}}
\def \vk{{\mathbf{k}}}
\def \vA{{\mathbf{A}}}
\def \vC{{\mathbf{C}}}
\def \vD{{\mathbf{D}}}
\def \vV{{\mathbf{V}}}
\def \vd{{\mathbf{d}}}
\def  \vSig{{\mathbf{\Sigma}}}
\def  \vsig{{\mathbf{\sigma}}}
\def  \vLa{{\mathbf{\Lambda}}}
\def  \veps{{\mathbf{\epsilon}}}
\def  \En{{\mathrm{E}}}
\def\bra{\langle}                                                                                                                                                                                                                                                                    
\def\ket{\rangle}
\def\wt{\widetilde}\begin{abstract} 
We have constructed the geometric phases emerging from the non trivial topology
of a space-dependent magnetic field $ \vB(\vr) $, interacting with the spin magnetic 
moment of a neutral particle. Our basic tool, adapted from a previous work on 
Berry's phases, is the space-dependent unitary transformation $\mathcal{U}(\vr)$ which leads to the identity,
 $ \mathcal{U}(\vr)^{\dag}\, \vS \cdot \vB(\vr) \,\mathcal{U}(\vr) = \vert \vB(\vr) \vert\, S_z$, at each point $\vr$. In the ``rotated'' Hamiltonian $\widehat {\mathcal H}$, $\frac{ \partial}{\partial \vr } $ is replaced by the non-Abelian covariant  derivative
$  \frac{ \partial}{\partial \vr }- \frac{i}{\hbar}\mathcal{A}(\vr)$
  where  $\mathcal{A}(\vr) = i \hbar \,\mathcal {U}^{\dag}.  \frac{ \partial}{\partial \vr }\mathcal{U}$ 
   can be written  as $\vA_1(\vr) S_x +\vA_2(\vr) S_y +\vA_3(\vr) S_z  $. The  Abelian differentials $\vA_k(\vr)\cdot d\vr$  are given in terms of the Euler angles  defining   the orientation of $\vB(\vr)$.
The non-Abelian field $\mathcal{A}(\vr)$ transforms as a Yang-Mills field, however its vanishing ``curvature''  reveals its purely geometric character.  We have defined a perturbation scheme based upon 
the assumption  that in $\widehat{\mathcal H}$  the longitudinal field $\vA_3(\vr) $ dominates the transverse field $\vA_{1,2}(\vr) $ contributions, evaluated to second-order. 
The geometry embedded in  both the  vector field $\vA_3(\vr) $ and the geometric magnetic field 
$\mathbf{ B}_3 (\vr) = \frac{ \partial}{\partial \vr }\wedge {\vA}_3(\vr)$
 is described by their associated Aharonov-Bohm phase.
 As an illustration we study the physics of cold $^{171}$Yb atoms dressed by overlaying two circularly-polarized stationary waves with orthogonal directions, which form a 2D square optical lattice. The frequency is tuned midway between the two hyperfine levels of the $(6s6p)^3P_1$ states to protect the optical $\vB(\vr)$ field generated by the lattice from the dressed atom instability. The geometric field $\vB_3(\vr)$ is computed analytically in terms of the Euler angles.  
The magnitude of the second-order corrections due to the transverse fields can be reduced to the percent level by a choice of light intensity which keeps the dressed atom loss rate $\leq 5$~s$^{-1}$. 
  A second optical lattice can be designed to confine the atoms inside 2D domains where $\vB_3(\vr)\cdot \hat z\geq 0$. 
 We extend our analysis to the case of a triangular lattice.  
\end{abstract}
 \maketitle

\section{Introduction }

In this paper  we shall be interested in the spatial geometry  associated with  the time-independent 
Hamiltonian    
\be
\mathcal{H}=\frac{\vp ^2}{ 2\, M} +V(\vr ) - \gamma_S \vS \cdot \vB(\vr),   
\label{H} 
\ee 
describing the quantum evolution of a non-relativistic neutral particle
of spin S and magnetic moment $ \gamma_S \; \vS $, interacting  with a scalar  potential
$ V(\vr )$ and a static non-uniform magnetic field $ \vB(\vr) $.
  
  In a recent work concerning Berry's phases \cite{Berry} generated by arbitrary spins non-linearly 
 coupled to time-dependent external fields \cite{bou1}, we have found that the geometry
 of the Hamiltonian  parameter space is more clearly exhibited if one uses a rotating frame  method instead of  the standard approach  involving space-time wave functions within the adiabatic  approximation. The Coriolis effect generates in the rotating frame  a time-dependent linear spin 
 coupling $ \Delta H(t) = -\gamma_S \Delta\vB(t).\vS $. The Berry phase is 
 generated by the longitudinal effective field  $ \Delta B_z (t) $ and the non-adiabatic corrections,  governed by the transverse fields  $ \Delta B_{x, y} (t)$,  are readily
 obtained  by a second-order perturbation calculation.  
 
 The purpose of this paper is to use a similar approach to construct the geometrical phase associated with the non-trivial spatial topology of the magnetic field $\vB(\vr)$. We are going to apply to $\mathcal{H}$ a local unitary transformation, which makes diagonal the spin coupling at each point $\vr$. The transformed kinetic term is obtained by replacing the gradient $\frac{ \partial}{\partial \vr }$ by the non-Abelian covariant derivative, $\vD  =  \frac{ \partial}{\partial \vr }-
     \frac{i}{\hbar} {\vA}_k (\vr)  S^k$ (with $S^1= S_x, \; S^2= S_y, \; S^3=S_z$), in which is encapsulated all the geometry of $\vB(\vr)$. By analogy with Berry's phase it is possible to assume, within well-defined conditions,  that the longitudinal field $\vA_3(\vr)$ gives the dominant contribution, the transverse ones, $\vA_{1,2}(\vr)$, leading to second-order corrections. This is a kind of Born-Oppenheimer  approximation, playing, here, the role of the adiabatic approximation. The role of the Berry phase is, then, taken up by the Aharonov-Bohm phase associated with $\vA_3(\vr)$. 
     
      A simple example of this problem, where the spatial dependence of $\vB(\vr)$ results from a constant field gradient, was studied experimentally by W. Phillips and coworkers \cite{Phil,Lin}. In the case where   $\vB(\vr)$ is a periodic field,  $\mathcal{H}$   could be 
 used, for instance as a model  to describe  the scattering of thermal  neutrons by  
 magnetic  materials \cite{Boni}. 
 There  has been  recently a regain of interest in this model for describing the evolution of cold atoms within an optical lattice 
 \cite{atom1,Coo,atom2,atom3}. ln this case, $\vB(\vr)$  stands for the   ``effective'' magnetic field  generated by an ac-Stark effect involving  the electric  field $\vE ( \vr ) $  associated with the coherent laser field   used to construct the optical lattice.  Two  realistic and simple examples  involving  $ S= 1/2$ cold atoms trapped in a two-dimensional (2D)  optical lattice will be  discussed in the last section of this paper.  
  
   \section{Non-Abelian gauge fields generated by the $\vB(\vr)$ geometry}
     \subsection{ The `` rotated frame" approach }
 We are going to extend the method we have used in the case of time-dependent fields to the  present context, by introducing   at each space point $\vr $ a  ``rotated frame". The change  of frame 
 is defined  by writing the  vector state as  $ \Psi(\vr) =\mathcal{U}(\vr) \widehat \Psi (\vr) $.
      The  space-dependent unitary transformation $\mathcal{U}(\vr) $ 
   is    designed to line up,  at each space point,  the magnetic field $\vB(\vr) $ 
   along the spin quantization axis:
    \be 
    \mathcal{U}^{\dag}(\vr)\, \vS \cdot \vB(\vr) \,\mathcal{U}(\vr) = \vert \vB(\vr)\vert\, S_z .
    \ee

  Let us ignore,  for a moment,  that  $\mathcal{U}(\vr) $  does not commute with the kinetic energy term. For  a periodic magnetic field,  
  one would then have to solve, for each spin component $m$,  
  a wave equation involving  a scalar potential, like 
in the historical experiment of Stern and Gerlach. One  would then come down to 
 solving standard problems of solid state theory.   
  
  To proceed, let us  introduce the rotation, 
  $ {\cal R}(\phi,\theta)= R(\hat{z}, \phi(\vr )) 
  \cdot R(\hat{y}, \theta(\vr ) ), $
  where $  R(\hat{n}, \chi )$  is the rotation of angle $ \chi $
   about the unit vector $\hat{n} $ and $\phi(\vr ), \;\theta(\vr )$ 
    the two Euler angles which define the direction of the $\vB(\vr)$ field.  
  More precisely, $\mathcal{R}(\phi(\vr ), \theta(\vr ))$  is  chosen in order  
   to satisfy the following equation, 
  \be   \vB(\vr)=   \vert \vB(\vr)\vert  \, \mathcal{R}(\phi(\vr ), \theta(\vr )) . \hat{z}. \ee
 We should keep in mind that there is not a unique way to bring $  \hat{z} $ along the direction of $\vB(\vr)$.
 This property can be associated with the gauge invariance of the non-Abelian gauge  fields that we are going to 
introduce.  
      It is easily verified that   $\mathcal{U}(\vr) $  given below  coincides with 
        the  unitary   transformation  associated  with the rotation  $ \mathcal{R}(\phi(\vr ), \theta(\vr )) $:
   \be 
  \mathcal{U}(\vr)  = \exp (\frac{- i }{\hbar}\, S_z\,\phi(\vr )).
   \exp (\frac{- i}{\hbar} \, S_y\,\theta(\vr )).
   \ee
Using group theory arguments,   one  can  derive the basic relation (see \cite{bou1}),  
\be
  \mathcal{U}^{\dag}(\vr)  \,\vS \,\mathcal{U}(\vr)=    
  \mathcal{R}(\phi(\vr ), \theta(\vr )) \cdot \vS . 
  \ee
This leads to the set of  identities:
 $ \mathcal{U}^{\dag}(\vr)  \,\vS\cdot \vB \,\mathcal{U}(\vr)=  ( \mathcal{R}(\phi, \theta) \, \vS) \cdot \vB= \vS \cdot (    \mathcal{R}^{-1}(\phi, \theta) \, \vB  )= \vert \vB(\vr)\vert\, S_z . $  
 
\subsection{ The geometric non-Abelian gauge fields}  
 Our next step is to write within  the rotated frame the eigenvalue equation, $ \mathcal{H} \,\Psi_n =E_n \Psi_n $,  now taking into account that $\mathcal{U}(\vr)$ does not commute with the kinetic energy operator $\frac{\vp ^2}{ 2\, M}  $.
  By making simple manipulations, we arrive at  
  the eigenvalue   equation written in  the rotated frame:
 $  \widehat{\mathcal{H}} \,\widehat\Psi_n(\vr)= E_n \widehat\Psi_n(\vr)$, 
 where the rotated Hamiltonian $\widehat{\mathcal{H}}=  \mathcal{U}^{\dag}(\vr)  \,  {\mathcal{H}} \;\mathcal{U}(\vr)$ is given by:
 \bea 
 \widehat{\mathcal{H}}&=& \frac{1}{ 2\, M} \widehat{\vp }^2+V(\vr )
  - \gamma _S \vert \vB(\vr)\vert\, S_z ,  \\
 \widehat{\vp }& =&  -i \hbar\frac{ \partial}{\partial \vr }
-i \hbar \,\mathcal {U}^{\dag}(\vr). ( \frac{ \partial}{\partial \vr } \mathcal{U}(\vr) ).
 \eea  
 In  reference  (\cite{bou1} sec. 4.2), we have made  a similar calculation 
 in the case where the Euler angles are  time- instead of space-dependent. The 
 needed result  is obtained   by doing the replacement:
  $ \dot{\phi} \rightarrow\frac{ \partial}{\partial \vr} \,\phi \, ,\,  
 \dot{\theta} \rightarrow \frac{ \partial}{\partial \vr} \,\theta \,,\, 
 \dot{\alpha} \rightarrow 0 $. This leads to the following expression  for 
 $\widehat{\vp }$:
 \bea
 \widehat{\vp }& =& -i \hbar\frac{ \partial}{\partial \vr }  - {\vA}_1 (\vr) S_x - {\vA}_2(\vr) S_y  -{\vA}_3 (\vr) S_z, 
 \nonumber \\
  {\vA}_1(\vr)&= &-\sin\theta(\vr)  \,\frac{ \partial}{\partial \vr }\phi(\vr),
 \nonumber \\
{\vA}_2(\vr)&= &\frac{ \partial}{\partial \vr} \theta(\vr), \;\;\;\;
 {\vA}_3(\vr)= \cos\theta(\vr)\,\frac{ \partial}{\partial \vr }\phi(\vr). 
 \label{objetA}
  \eea
  Note that $\vA_3(\vr) \, S_z$, {\it etc.} are tensor products of a vector operator in the geometric space by a vector operator in the spin space.  A  remarkable  feature of the above results  is also the fact that the three fields ${\vA}_i (\vr)$  do not depend on the value of  $\vS^2=S(S+1) \hbar^2$.  This  follows 
 from the fact that  the evaluation of  $\widehat{\vp }$ relies  only upon the $SU(2)$  Lie algebra: $[ S^i,S ^j]=i\,\hbar \, {\epsilon}_{i\,j \,k} \, S^k $.  A  simple  direct evaluation of the above  formulas  has been performed for $S= 1/2$  using the fact that    $\mathcal{U}(\vr) $ is given  by a $2\times 2$ matrix, linear with respect to $\vS$.
From now on,   we shall  often use the notation $ S^1=S_x, \; S^2=S_y, \; S^3=S_z $ in order to stress that the  $S^i $ will be treated   as scalar objects under  ordinary space  rotation.  It is of interest to incorporate the three fields ${\vA}_i(\vr)$ into a   ``non-Abelian"  classical field $\mathcal{A}(\vr) $, a well-known concept used in other fields of physics:
\be
 \mathcal{A}(\vr) =  {\vA}_k (\vr)  S^k = i \hbar \,\mathcal {U}^{\dag}(\vr). ( \frac{ \partial}{\partial \vr } \mathcal{U}(\vr) ) .
 \label{noAbA}
 \ee
  The Hamiltonians considered in this work  have not any  simple  invariance  properties  under rotation involving  the total angular momentum of our neutral particle
  $ \vJ= \vL  +\vS $. Using a fashionable expression, our spin will be treated as a ``colored" spin.
    \subsection{ Non-Abelian covariant derivatives and gauge invariance}
  In analogy with  what is done for Abelian fields,  we define the   basic concept of  
   covariant ``derivative"  as  follows:
  \be
  \vD =  \frac{i}{\hbar}  \widehat{\vp } =  \frac{ \partial}{\partial \vr }-
     \frac{i}{\hbar} {\vA}_k (\vr)  S^k = 
   \frac{ \partial}{\partial \vr } -\frac{i}{\hbar}  \mathcal{A}(\vr) .
  \label{covD}
   \ee
  We are now  going to require that   $ \vD. \widehat{\Psi}(\vr)$  transforms under  any space-dependent $ SU(2)$   unitary transformation 
   $ \exp (\frac{i}{\hbar} \Lambda_{k}(\vr)\, S^k  )$  in the same way as the wave function  $\widehat {\Psi}(\vr)$, {\it i.e.}
$ \widehat{\Psi}(\vr) \rightarrow \widehat {\Psi}_{\Lambda}(\vr) =
          \exp (\frac{i}{\hbar} \Lambda_{k}(\vr) S^k  ) \, \widehat {\Psi}(\vr) ).$
  The  above  requirement  will  be 
    satisfied  provided  that the non-Abelian field        $\mathcal{A}(\vr)$  is subjected to a well-defined  non-Abelian   gauge transformation: 
     $   \mathcal{A}(\vr)\rightarrow    \mathcal{A}_{\Lambda}(\vr)$.  
 The covariance condition  means  that   
 the operator $\vD$  should  transform locally  as a vector operator under any space-dependent $ SU(2)$   unitary transformation: 
   \be  
   \frac{ \partial}{\partial \vr }-
  \frac{ i }{\hbar}  \mathcal{A}_{\Lambda} 
   = \exp (\frac{i}{\hbar} \Lambda_{k} S^k  ) \,\vD\,
      \exp (-\frac{i}{\hbar} \Lambda_{k} S^k  ). \ee 
      
The   non-Abelian gauge transformation  $   \mathcal{A}(\vr)\rightarrow    \mathcal{A}_{\Lambda}(\vr)$    is then  easily  derived by introducing in  the above formula the expression of  $\vD $ given in Eq.(\ref{covD}), 
    \bea     
\mathcal{A}_{\Lambda}(\vr) &= &  \exp (\frac{i}{\hbar} \Lambda_{k} S^k  )
 \,\Big (\frac{ i\,\hbar \partial}{\partial \vr } \exp (-\frac{i}{\hbar} \Lambda_{k} S^k  )
 \Big ) \nonumber\\
 & &+  \exp (\frac{i}{\hbar} \Lambda_{k} S^k  ) \,\mathcal{A}(\vr)\,   \exp (-\frac{i}{\hbar} \Lambda_{k} S^k  ).
 \label{GaugeTransf}
\eea
  It is found to  agree with the formulas  given  by S. Weinberg \cite{Wein} for general non-Abelian  fields. If the space-dependent  matrix
  $ \Lambda_{k}(\vr) S^k $ is replaced by an ordinary scalar function of $\vr$, one recovers the usual Abelian gauge transformation. 
  
  An important physical concept for non-Abelian gauge  fields is the ``curvature" tensor,
  involving  covariant derivative commutators, $ \mathcal{F}_{i \,j}(\vr )=i [  D_i,  D_j ]$, or
  $$\mathcal{F}_{i \,j}(\vr ) =
  \frac{ \partial}{\partial x_i }\mathcal{A}_{x_j} (\vr)
  - \frac{ \partial}{\partial x_j }\mathcal{A}_{x_i} (\vr) -i [ \mathcal{A}_{x_i},\mathcal{A}_{x_j} ] .$$
 A crucial point, here, is that the non-Abelian field $ \mathcal{A}(\vr)   $ given  by equation (\ref{GaugeTransf}) can be expressed under two forms, which are equivalent  as a result of $\mathcal{U}(\vr) $ unitarity:
  $   \mathcal{A}(\vr) = i \hbar \,\mathcal {U}^{\dag}(\vr)\,. \,( \frac{ \partial}{\partial \vr } \mathcal{U}(\vr) )=   -i \hbar(  \frac{ \partial}{\partial \vr } \mathcal{U}^{\dag}(\vr) )\,  . \,
 \mathcal{U}(\vr) ) $. Using this identity to rewrite  $\mathcal{F}_{i \,j}(\vr ) $ one arrives, after some algebraic manipulations,
 at the remarkable result: $ \mathcal{F}_{i \,j}(\vr )=0 $.  This shows, clearly, that
 \emph {  $\mathcal{A}(\vr)  $} is not  a self interacting Yang-Mills  field like the fields 
  used in Particle   Physics, but a purely geometrical field
 associated with a local change of space coordinates. There is  a similarity
  with the situation encountered in the  tests of  the ``equivalence principle"  which involve  ``flat"  space-time metrics.  
     
        \begin{figure*}
 \includegraphics[height=10.0 cm]{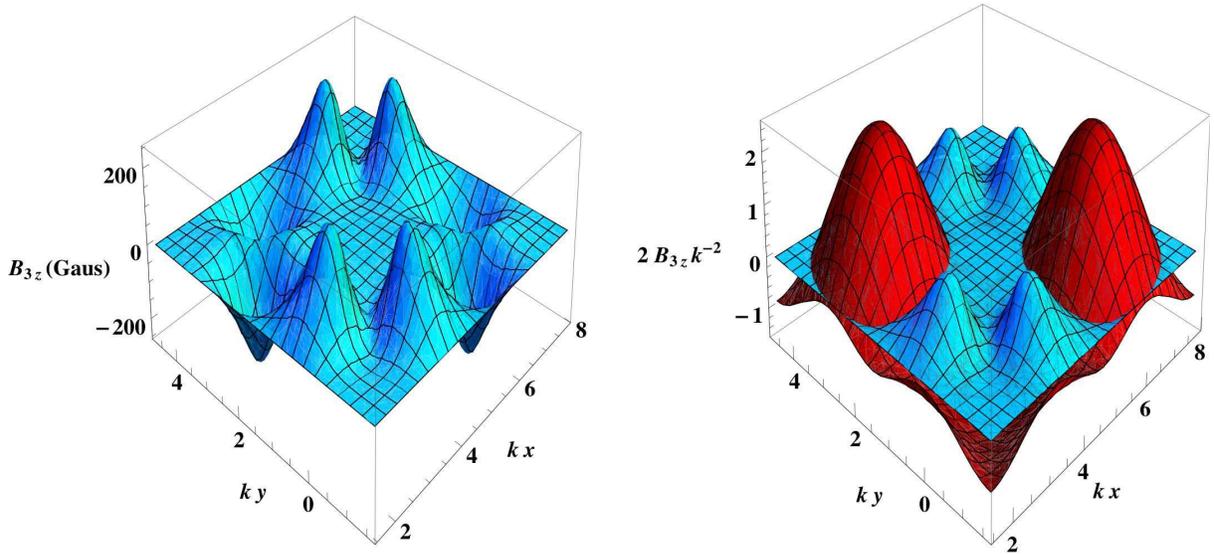}
\caption{ The geometric magnetic field $ B_{3z}$ and the confinement potential inside one elementary cell of the optical lattice represented in the ``rotated frame''. The {\it Left} graphic gives a perspective 3D plot  of $\vB_{em}(\vr) $, the electromagnetic field equivalent  of the geometric field $\vB_3(\vr)=
 \frac{ \partial}{\partial \vr }\wedge {\vA}_3(\vr)$,  as a function of   $ k \,x $  and $ k \,y $  ($k$ being the wave number of the optical ``dressing" field generating  the non-Abelian  gauge fields). The geometric magnetic field $ B_{3z}$   exhibits  two positive maxima  within the plaquette  
defined by   $ \pi/2 \leq   k\,x \leq 3\pi/2  $, $-\pi/2  \leq  k \, y \leq  \pi/2 $. 
   A  symmetric  pair of negative minima $ B_{3z}$  appears within the plaquette    
  obtained by performing the translation $ x\rightarrow  x+\pi $,  as predicted by
  the formula (\ref{Bgeom}). 
In the {\it Right}  graphic we have plotted   $ B_{3z}$ together with 
  the total potential   $ V_c(\vr )=\mathcal{V}(\vr)  +  \mathcal{W}(\vr) - 
\gamma_I \, \hbar/2 \,\vert\vB (\vr) \vert $  
 acting upon a state  with $ I_z = \hbar /2 $,  drawn in dark red using as unit $\gamma_I \, \hbar \,B_0 $.
 ( For the sake of visibility $ V_c(\vr )$  has been shifted up by one unit.)
 As expected,  the  added  potential $\mathcal{W}(\vr) $   does  prevent the cold atoms from exploring the
  regions where  the sign of $ B_{3z}$ is negative. Indeed,  the   pair 
  of negative minima $ B_{3z}$   of the {\it Left} are clearly not accessible for atoms below a 
  well defined temperature.
   It is of interest to note that the two  graphics   are  invariant
 under  a  rotation   of  $\pi $ around  a vertical
 axis  centered at  the point   $ k x_0 = 3\pi/2, \, \,k y_0=  \pi/2  $,  leading to the coordinate transformation, 
 $ x  \rightarrow   x' =3\pi -x , y  \rightarrow   y' =\pi -y $   which leaves   invariant
 both   $ B_{3z} (\vr ) $ and $\mathcal{W} (\vr). $  The values  of these quantities for the
 full 2D space are  obtained  by performing  the  set of translations:
 $ x  \rightarrow   x' =2 \,n_1 \,\pi  + x,  \,  y  \rightarrow   y' =  2\, n_2 \,\pi  + y $ where
 $  n_1$ and  $  n_2 $ are arbitrary integers.}
 \label{Fig1}
\end{figure*}
A possible application of the non-Abelian gauge transformation   
    is  the choice of the Landau  gauge for ${\vA}_{\Lambda 3} (\vr)$ by writing 
      $ { A_{\Lambda 3 x}=\hat {x}\cdot\vA}_{\Lambda 3} (\vr) =0 $  without affecting the magnetic coupling.  
  It is  easily seen from  Eq. (11),  that the wanted gauge transformation  has to
  satisfy the following constraints:
$
\widetilde{\Lambda_{1}}= \widetilde{\Lambda_{2}}=0  \; , \; 
\frac{\partial}{\partial x }\, \widetilde{\Lambda_{3}}(\vr)= -{A}_{3 x} (\vr) .
$
An explicit expression $ \widetilde{\Lambda_{3}}(\vr) $  is obtained by  a  quadrature over 
$y : 
  \widetilde{\Lambda_{3}}(\vr)=- \int_0^x A_{3 x} (v,y,z) \, dv. $
   With this choice of $\Lambda$ the Hamiltonian 
  $\widetilde{\mathcal{H}}$  is obtained from $\widehat{\mathcal{H}} $ by
  performing the replacement ${\vA}_i(\vr) \rightarrow   {\widetilde \vA}_i(\vr)   $,  
   \bea   
    {\widetilde \vA}_1(\vr)    &= & \cos( \widetilde{\Lambda_{3}}) {\vA}_1(\vr) + 
    \sin( \widetilde{\Lambda{3}}) {\vA}_2(\vr),   \nonumber\\
    {\widetilde \vA}_2(\vr)    &= & - \sin( \widetilde{\Lambda_{3}}) {\vA}_1(\vr)+
    \cos( \widetilde{\Lambda_{3}}) {\vA}_2(\vr) ,\nonumber\\
    \widetilde {\vA}_3(\vr) &= &( A_{3 y}(\vr ) + \frac{ \partial}{\partial y }
     \widetilde{\Lambda_{3}}(\vr) )\,\hat{y}.
     \label{Atild}
   \eea
    \section{   The  Aharonov-Bohm phase from the ``longitudinal" gauge field  } 
 Although the above developments  are valid for any spin values, from now on, for sake of simplicity, {\it we  shall limit ourselves to the case of a $ S=\frac{1}{2} $  particle in a 2D space.} So far we have made no approximation.  
In analogy  with  the  method we have used  for Berry's phases generated by non-linear
 spin Hamiltonians \cite{bou1}, we 
  are going in sec. IV to construct a perturbation   
 expansion, using as starting point the diagonal part  
  $ {\widehat{\mathcal{H}}}_0 $  of the   Hamiltonian $\widehat{\mathcal{H}}$, 
\bea    
{\widehat{\mathcal{H}}}_0  &  =   & - \frac{1}{2M} \big(\frac{ \hbar \partial}{\partial \vr }\big)^2 + V(\vr )+  \frac{\hbar^2}{8\, M}  \,\sum_{i =1}^{i=3} \vA_{i} (\vr) ^2   \nonumber \\ 
& &   - \big(\frac{ i\hbar}{ 2\, M}
 \{ \frac{ \partial}{   \partial  \vr }, {{\vA}}_{3} (\vr) \} 
   +\gamma_S \vert \vB(\vr)\vert\, \big)\, S_z ,  
   \label{Hhat0}
\eea
$\{...\}$ denoting the anticommutator between operators.
The above approximation is often referred to as a  Born-Oppenheimer (BO) approximation. 
It supposes that the evolution of the spin wave function is slow as compared to that of the spatial wave function describing the particle internal state (just like in molecular physics, the motion of the nuclei of a molecule is generally slow in comparison with that of the atomic electrons).  The validity of this approximation in the present context is dicussed later on (see sec. IV).
This is one advantage of the present  ``rotated" frame approach, similar to our treatment   
of the adiabatic approximation \cite{bou1}, to  provide a systematic perturbation method to calculate the corrections.
The left over perturbation Hamiltonian is given by the non-diagonal contribution of ${\widehat{\mathcal{H}}}$,  
\be
 {\widehat{\mathcal{H}}}_1=  -  \frac{i \hbar}{ 2\, M}  \frac{ \partial}{\partial \vr } 
 \cdot \big ( \vA_1 (\vr) S_x + \vA_2 (\vr) S_y \big) + h.c. 
 \label{Hhat1}
 \ee
  
   Let us discuss the  physics content of Eq. \ref{Hhat0} for a given  eigenvalue of 
    $   S_z/ \hbar,\; m=\pm  \frac{1}{2}$.
   The three  first terms  describe a  spinless particle moving in a scalar potential.
   The  terms of the second line represent the coupling of this  particle  of effective electric charge $m$,  interacting with  the  Coulomb-like  potential $ -\gamma_S \vert \vB(\vr)\vert $   and  a magnetic vector 
   potential given by  $\vA_3(\vr). $  The diamagnetic term ${\vA}_{3} (\vr) ^2 $ is incorporated into  the third term of the  first line.
   
   Ignoring for a moment ${\widehat{\mathcal{H}}}_1$, we would like  to consider  a two-slit
   Aharonov-Bohm interferometer experiment involving  two beams described by wave packets,  
   solutions of the Schr\"odinger equation associated with  ${\widetilde{\mathcal{H}}}_0$.  It is  convenient
   to  describe the two  interfering beams  by two Feynman path integrals associated with the  
   classical paths $\mathcal{C}_1$ and  $\mathcal{C}_2$  enclosing  a 2D  finite region   
     $\mathcal{R}_B $ where the ``geometric   magnetic" field is \emph{confined}:
     \bea
        \mathbf{ B}_3 (\vr) &= &\frac{ \partial}{\partial \vr }\wedge {\vA}_3(\vr) =
      \sin(\theta(\vr) ) \Big ( \frac{ \partial}{\partial x }\phi(\vr) \, \frac{ \partial}{\partial y }\theta(\vr) -
        \nonumber \\
      && \frac{\partial}{\partial y }\phi(\vr) \, \frac{ \partial}{\partial x  }\theta(\vr) \Big ) \,  \hat z. 
      \label{B3}
        \eea
           Following Feynman \cite{Feyn},  the phase difference
         \be 
           \Phi_{A B}(m) =   - \,m \Big (  \oint_{ \mathcal{C}_1 }\, {\vA}_3(\vr_1) \,  d{\vr}_1 
          -\oint_{ \mathcal{C}_2 }\, {\vA}_3(\vr_2) \,  d{\vr}_2 \Big ), 
          \label{ABphase}
          \ee
               is factored out of  the functional integration and  
    is given explicitly by the 2D integral:
    \be
     \Phi_{A B}(m)=m\int_{\mathcal{R}_B}   dx \,dy  \,B_{3 z }(x,y),  
     \label{phiAB}
    \ee
    where $m \hbar$ plays the role of the electric charge.
     The Aharonov-Bohm  phase \cite{ABphase} $\Phi_{A B}(m) $, can be viewed as  a  spatial extension 
     of  Berry's phase associated with the non-trivial geometry  of the  \emph{physical} $\vB(\vr) $  field, which is described  by the \emph{geometric}  magnetic field, $\mathbf{B}_3(\vr)$. 
    
     The above calculation has been performed in the rotated frame.  To return to the 
     laboratory frame, we shall use the fact that, in a typical Aharonov-Bohm   experiment, 
     the paths  $  \mathcal{C}_1$ and $\mathcal{C}_2    $  are  drawn in  regions where 
  $\mathbf{ B}_3 (\vr) =0$.  Assuming    that  the  Feynman phase factors  $ \oint_{ \mathcal{C}_i }\, {\vA}_3(\vr_i) \,  d{\vr}_i $ have been factored out, 
  the wave functions relative to the    paths  $\mathcal{C}_i $ 
  can be written in the rotated frame as 
  $  \widehat{ \Psi}_{i,m} = \widehat{ \psi}_{i,m} (\vr,t) \otimes \widehat{\chi}_m $.  The spatial wave functions $\widehat{ \psi}_{i,m} (\vr,  t) $  obey the Shr$\ddot{o}$dinger equation  associated with the 
 Hamiltonian $ {\widehat{\mathcal{H}}}_0 (\vA_3 =0, S_z =m\hbar )$ given
 by Eq. (\ref{Hhat0}); $\chi_m$ is the $ S=1/2 $ spinor wave function, 
  $ ( \delta_{m,1/2}, \delta_{m,-1/2})$. When going back to the laboratory 
  frame  the spatial wave function $\widehat{ \psi}_{i,m} (\vr, t) $ is unchanged, whatever $\vr, t$, since $\mathcal{U}(\vr, t)$ acts only on the spin, while the spinor, 
   ${\chi}_{i,m}(\vr )= \mathcal{U}(\vr, t) . \, \widehat{ \chi}_{i,m} $ has been rotated. However, at the 
   interference point the spinor wave function interference in the laboratory and in the rotated frame is identical, $\bra \widehat{ \chi}_{2,m} \vert \widehat{ \chi}_{1,m} \ket= \bra\chi_{2,m}\vert \mathcal{U}^{\dagger} (\vr) .\, \mathcal{U}(\vr)\vert\chi_{1,m}\ket$, as a consequence 
   of the unitarity relation $ \mathcal{U}^{\dagger} (\vr) .\, \mathcal{U}(\vr)=1 $.
   In more physical terms the spins rotate differently along the two paths 
   $  \mathcal{C}_1$ and $\mathcal{C}_2    $, but their  scalar product is preserved at the interference point. In conclusion:  $\Phi_{A B}^{lab}(m) = \Phi_{A B}(m)$.
   
         In order to clarify the relation of $\Phi_{A B}(m) $    with the standard Berry's phase, it is instructive to discuss briefly the case of a  space and time-dependent $\vB(\vr,\, t)$ field.
      As before,  let  us  construct  
  the wave function in the frame rotating in space-time:
    $ \widehat\Psi (t)=\mathcal{U}^{\dagger} (\vr,t) \Psi (t)  $,  where  $\Psi (t) $ is a
    solution of  the Schr\"odinger equation associated with the time-dependent  Hamiltonian
 $ \mathcal{H}(t) $  obtained by giving a space- and time-dependence to
 the magnetic field $\vB $. The unitary transformation
  $  \mathcal{U}(\vr,t)  = \exp (\frac{- i }{\hbar}\, S_z\,\phi(\vr,t )).  \exp (\frac{- i}{\hbar} \, S_y\,\theta(\vr,t )) $  brings the $\vB(\vr,t)$ field along the  $z$ axis.  The time evolution  of  the ``rotated" wave function is  then governed by the  Hamiltonian 
 $  \widehat{\mathcal{H}}(t)=  
 \frac{1}{ 2 M} (-i \hbar\frac{ \partial}{\partial \vr } - {\vA}_k (\vr,t)  S^k )^2 -
   {V}_k (\vr,t)  S^k + V(\vr )  - \gamma _S \vert \vB(\vr,t)\vert\, S_z $.  The scalar fields
    $ {V}_k (\vr,t) $ are obtained  from the vector fields $  {\vA}_k (\vr,t)$ by replacing in Eq.~(\ref{objetA}) $ \frac{ \partial}{\partial \vr } $ by 
    $\frac{ \partial}{\partial t } $. For the moment, let us   remove the $ \vr$  dependence    and ignore the transverse  potentials $ V_{1,2}(t) $,  governing  
    the  non-adiabatic corrections  \cite{bou1}. Berry's phase  is  closely related  to the  dynamical phase contribution     generated by  $ -V_3 (t) S_z$  in the rotating frame \cite{bou1}:
$ \widehat{\Phi}_{D 3} = \int_0^T dt\, m \cos\theta(t) \, \dot{\phi}(t) $. By returning 
to the laboratory frame, $ \widehat\Psi (t) \rightarrow \Psi (t)  $,  one recovers the standard Berry's phase formula: $\beta(m)=\int_0^T dt\, m \big( \cos\theta(t) - 1  \big)\dot{\phi}(t). $

The creation of effective magnetic fields in 2D optical lattices was suggested by Jaksch and Zoller \cite{Zol03}.   They deal with   a pseudo spin $\frac{1}{2}$   associated  with two internal atomic states. The effective fields are resulting from the spatial dependence of the Rabi frequencies induced by Raman lasers designed to transfer an atom  in a  given internal state to  a  nearest-neighbor  lattice site associated with a flipped internal state. In addition, there is in the Hamiltonian a dynamical contribution which specifies the direction of the atomic jump. This requires, in practice, additional space-dependent potentials \cite{Fgjd}. Here, we have adopted   a totally different point of view: we have focused upon  continuous  space-dependent  ``dressed" Hamiltonians, having  a linear  spin coupling. We have developed a precise formalism showing how   the non-trivial spatial topology of our  spin Hamiltonian implies the existence  of an  Aharonov-Bohm phase,  generated by  a geometric   magnetic field   $\mathbf{ B}_3 (\vr)$. The role of the electric charge is taken  up by the eigenvalue of $ S_z/\hbar, \, m  =\pm \frac{1}{2}$.
\section{The transverse gauge fields corrections}
We are going to present    a  brief    analysis of the effects or the transverse
gauge fields  ${\vA}_{1,2} (\vr)$ upon  the physics associated
with the approximation $ \widehat{\mathcal{H}} \approx {\widehat{\mathcal{H}}}_0 $, 
taking as an example the calculation of the Aharonov-Bohm phase. 
  We shall use a  perturbation approach using as starting point the eigenstates
 of ${\widehat{\mathcal{H}}}_0$ associated with a given eigenvalue of $ m $:
$ {\widehat{\mathcal{H}}}_0 \Psi  _{n\,m} (\vr ) =E_{n\,m}  \Psi _{n\,m} (\vr )$.
The  eigenenergy  corrections appear
  to  be of second-order with respect to ${\widehat{\mathcal{H}}}_1 $:
$\Delta_2 E_{n\,m}   = \sum _{\bar{n}}
   \vert \bra \Psi _{n,m} \vert {\widehat{\mathcal{H}}}_1
 \vert\Psi _{\bar{n},-m} \ket \vert ^2 \, / (E_{n,m}-E_{\bar{n}, -m} ) $.
 Assuming  that the    magnetic coupling  energy, $   -\gamma_S   S_z\vert \vB(\vr)\vert $, 
 is  dominant in    $ {\widehat{\mathcal{H}}}_0 $ allows us to write:
$ E_{n,m} -E_{\bar{n}, -m  }$  as the quantum   average of
$    -2 m\, \gamma_S  \, \vert \vB(\vr)\vert   $.
 The  energy shift
 $ \Delta_2 E_{n\,m} $   is then
 approximated by the rather simple expression:
 \be
\Delta_2 E_{n\,m}
  \approx  -  \frac{\bra \Psi _{n,m} \vert {\widehat{\mathcal{H}}}_1^2 \vert\Psi _{n,m} \ket}
   {    2\, m\, \hbar\gamma_S \,
   \bra \, \Psi  _{n\,m}\vert\,  ( \vert\vB(\vr)\vert ) \, \vert\Psi  _{n\,m}\ket  }. 
 \label{DeltaE}
 \ee

  Let us now turn to  the $ {\widehat{\mathcal{H}}}_1 $ corrections
  to  the geometric phase  $\Phi_{A B}(m)$.   In a typical atomic interferometry experiment, the convergence of the   two paths $ \mathcal{C}_{i}$ is implemented
  with a localized ``mirror" device. If the gauge vector field line integral is
  factored out,   the wave packets  $\Phi_i( \vr,t) $ associated  with  $ \mathcal{C}_{i}$
  satisfy, outside the ``reflexion" regions, the Schr\"odinger equation governed by
  ${\widehat{\mathcal{H}}}_0 ({\vA}_3=0)$,   a real operator having real eigenfunctions  $\Phi_{n,m} $ with eigenvalues ${\cal E}_{n,m}$.
  Let us  concentrate upon wave packets relative to the converging sections of the  paths,
 $ \Phi_{i}^{c} (\vr , t) $  with $ 0 <  t    < T $, where
  the times  $ -T, 0, T $  are respectively the emission, reflexion  and  interference times. At these three times occur  transformations described within the sudden approximation.  Using standard  time-dependent perturbation  formalism,
 the first-order  $ {\widehat{\mathcal{H}}}_1 $ corrections  to the converging wave
 packets reads:
 \bea
 \delta \Phi_i^c(t) &=& 2 \,i \sum_{n, \bar{n}}
  \bra\Phi_{\bar{n},-m}\vert{\widehat{\mathcal{H}}}_1\vert \Phi_{n,m} \ket
  \bra\Phi_{n,m} \vert\Phi_{i}^c (0)\ket  \nonumber \\
     && \times \; \frac{ \sin(  ({\cal E}_{n,m}- {\cal E}_{\bar{n},-m}) t/\hbar )} {( {\cal E}_{n,m}- {\cal E}_{\bar{n},-m})}\,\,
     \Phi_{\bar{n},-m}.
    \nonumber
 \eea
 Since  only the wave packets  with identical  $m$ can interfere,  the lowest-order interference  correction reads:
$   \bra \;\delta\Psi_{2}^c (T)\,\vert \,\delta \Psi_{ 1} ^c(T) \; \ket=
\bra \; \delta \Phi_{2}^c (T)  \,\vert \,\delta \Phi_{1}^c (T) \; \ket  \,\exp( i \Delta  \Phi_{AB }^c ) $,
where  $\Delta \Phi_{AB}^c$ is  the  phase difference accumulated   along the converging paths.
$ \Delta \Phi_{AB}^c=  m \big (  \oint_{ \mathcal{C}_{1}} \, {\vA}_3(\vr_1) \,  d{\vr}_1   -  (\mathcal{C}_{1} \rightarrow \mathcal{C}_{2})  \big  ).$ (Note the change of sign with respect to $ \Phi_{AB}$.)
 To proceed  we make two  approximations. As before in the evaluation
 of $  \Delta_2 E_{n\,m} $ we shall write:
 $ ( \mathcal{E}_{n,m}-  \mathcal{E}_{\bar{n} ,-m}  ) ={    2\, m\, \gamma_S \,
   \bra \, \Psi  _{n\,m}\vert\,  ( \vert\vB(\vr)\vert ) \, \vert\Psi  _{n\,m}\ket  } $.
 Then,
 we shall  drop the oscillating  $\sin$ products and replace the  $\sin^2$ by their average $1/2$.
 This leads to the  following expressions 
   for  $\delta_{1,2}^c=\bra  \delta \Phi_{2}^c (T)  \vert \delta \Phi_{1}^c (T) \ket$ and $\eta_{n,m} $, 
     \bea
    \delta_{1,2}^c  &    \approx  &
  \sum_{n}  \eta_{n,m} \; \bra \Phi_{2}^c (0)\vert \Phi_{n,m}\ket \bra\Phi_{n,m}
  \vert\Phi_{1}^c (0)\ket,   \nonumber \\
\eta_{n,m} &=&   \frac{2 \bra  \Phi_{n,m}\vert{\widehat{\mathcal{H}}}_1^2 \vert
  \Phi_{n,m}\ket  }  {  (  \gamma_S \, \hbar
   \bra \, \Psi  _{n\,m}\vert\,  ( \vert\vB(\vr)\vert ) \, \vert\Psi  _{n\,m}\ket )^2 },
    \eea
 which are closely related in magnitude to the 
 energy shift $\Delta_2 E_{n\,m} $, calculated previously (Eq. \ref{DeltaE}).

  It is  of interest to  give an   explicit expression of
$\bra  \Phi_{n,m}\vert {\widehat{\mathcal{H}}}_1^2 \vert   \Phi_{n,m}\ket $, 
 by writing $ {\widehat{\mathcal{H}}}_1  =
 \sum_{k=1}^{2}   {\widehat{\mathcal{H}}}_{1,k} \, S^k $
  where $ {\widehat{\mathcal{H}}}_{1,k} = \{  \vp \ , \,\vA_{k} (\vr) \} /(2 \, M).$
 Using the  $  S^k $ algebra for $ S=1/2$,  one gets the more compact expression, 
 
  $\bra  \Phi_{n,m}\vert {\widehat{\mathcal{H}}}_1^2 \vert   \Phi_{n,m}\ket =
  \sum_{k=1}^{2}  \bra  \Phi_{n,m}\vert {\widehat{\mathcal{H}}}_{1,k} ^2\vert   \Phi_{n,m}\ket   /4.$
  
  \noindent  Pulling    out   the $ \vA_{k} (\vr) $  field from the matrix 
   element of ${\widehat{\mathcal{H}}}_{1,k} $    one gets a formula useful  for an  explicit evaluation:
\bea
 \hspace{-2mm}\bra  \Phi_{n,m}\vert {\widehat{\mathcal{H}}}_1^2 \vert   \Phi_{n,m}\ket & = &
 \frac{\hbar^2}{4 \,M^2} \Big\{\sum_k \int d^3\vr \, \vert \phi_{n,m}\vert ^2  (\frac{ \partial}{  2 \,\partial  \vr  }\cdot \vA_{k} )^2   \nonumber \\
&&\hspace{-20mm} + \sum_{k,i,j}    \int d^3\vr
 \frac{ \partial}{ \partial  x_i } \Phi_{n,m}^*  \,\frac{ \partial}{ \partial  x_j }  \Phi_{n,m} \times
A_{ k,i}\,A_{ k,j} \Big\}
\label{H1sqav}
\eea

   Using the above  results which make \emph{no assumption} about the origin of $\vB(\vr)$, it is possible to give an order of magnitude of $\eta_{n,m}$ in the particular case of 
a periodic  field  $ \vB(\vr) $  associated with a single   wave number $\vk$.
 Looking at  Eq.~(\ref{objetA})  one sees clearly that the non-abelian  fields
  $\vA_i(\vr) $ scale  as $  \vk  $ and their divergence, 
   $ \frac{\partial}{  \partial  \vr  } \cdot\vA_{i}( \vr) $,  as $ k^2$.
 In an  explicit  computation to be presented later on, it appears that the
 terms involving the field divergence is the dominant one.
  Assuming that  the magnetic coupling is the dominant term in ${\widehat{\mathcal{H}}}_0$, one arrives at the estimate, 
   \be
 \hspace{-1.mm}\vert\eta_{n,m} \vert \lesssim \eta_{max} \;\;\; {\rm with} \;\; \eta_{max}=\frac{1}{2} \Big ( \frac{ (\hbar \vk)^2 /(2 \,M ) }{\hbar \gamma_S  { \vert \vB(\vr)\vert}_{min}} \Big ) ^2 , 
   \label{eta}
    \ee
 which can also be viewed as a scaling law. 
 In the case where the periodic field $ \vB( \vr ) $ contains an effective optical magnetic
  field $ \vB_{opt}( \vr ) $    generated inside a 2D optical lattice, 
  $\ (\hbar \vk)^2 /(2\,M)  $ is  the recoil energy of the cold
   atom following the absorption of one photon of momentum $\vk$.
   \section{Application to $^{171}\rm {Yb}$ cold  atoms in a 2D optical lattice}
We now want to illustrate the above theoretical analysis with  a simple realistic example. We choose the case of cold neutral atoms with  $F=I=\frac {1}{2}$ angular momentum,  ``dressed'' by a second-order Stark effect \cite{cct} involving the wave electric field $\vE(\vr)$ of an optical lattice.  The light wave directions and polarizations are selected to generate an effective magnetic field $\vB_{opt}(\vr)$ which has the lattice periodicity.

Much more ambitious experimental programs involving 2D optical lattices are aiming at a simulation of actual condensed matter problems \cite{atom4,atom5,Fgjd,atom6}, like the electronic properties of graphene.
 \subsection{The  dressed atom Hamiltonian $ \mathcal{H} $}
Let us assume  that the  ``dressing" optical electric field $ \vE(\vr) $ is obtained  by  the coherent superposition of two  stationary waves, directed along $\hat x$ and $\hat y$.  Each one results from the interference of a circularly polarized laser beam (wave vector $\vk$, helicity $\xi$) and of the beam reflected at normal incidence, all beams carrying along their propagation axis the same angular momentum $\hbar \xi$ per photon,
\bea
\vE(\vr) \hspace{-2mm} &=&\hspace{-2mm}\mathcal {E} \Big\{\frac{1}{\sqrt 2}(\hat y + i \xi \hat z)  ( \exp{i k\,x} - \exp{-i k\,x})\nonumber  \\
   &&\hspace{-2mm}+\frac{1}{\sqrt 2}(\hat z + i \xi \hat x) (\exp{i k\,y}  - \exp{-i k\,y})  \Big\}. 
   \label{statwave}
  \eea
 This configuration has been selected in order to make the space dependence of $ i\vE ^{*}\wedge \vE $ appearing in $\mathbf B(\vr)$ both simple and symmetric under  the $x\rightarrow y$ exchange (see Eq. (\ref{Bopt}) below).
 In particular the relative phase of the two stationary waves plays an essential role. 
    To avoid the  singularities of   the non-Abelian
 gauge fields  directly associated with the  geometry of
 the  optical  ``effective"  magnetic field $ \vB_{opt} $,
 a  uniform dc magnetic field $\vB_0$ has to be  applied, with  the same
  $x \leftrightarrow y $ symmetry as $ \vB_{opt} $, 
$ \vB_0=  ( \sin{u} \; (\hat x+\hat y)/\sqrt{2} + \cos {u} \;\hat z ) B_0 $.
 
 We consider non-interacting $^{171}$Yb atoms, with nuclear spin $\frac{1}{2}$, trapped in this 2D optical lattice and ignore many-body effects. The laser frequency is detuned from the transition at 555.8 nm between the ground state $(4f^{14}, 6s^2)^1S_0$ and the first $(4f^{14}, 6s 6p)^3P_1$ excited state.
In the case of a single transition between two states of angular momentum $\frac{1}{2}$, the optical field generates a space-dependent effective dc magnetic field $ \vB_{opt}$ which
follows from the identity, valid for $ S=1/2$, 
 \be
 (\vS \cdot \vE ^{*})(\vS \cdot \vE ) = \frac{\hbar^2}{4} \vE ^{*}\cdot \vE 
+i \frac{\hbar}{2}  \vE ^{*}   \wedge  \vE\cdot  \vS. 
\label{Sid}
\ee
 The  magnitude  of   $ \vB_{opt}$ scales as the atom-field coupling $ \Omega^2/\delta$, where $\hbar \Omega = 2 \,\mathcal{E}\, d $ is the Rabi frequency of the transition, 
$d= \langle ^3P_0, \; m=0 \vert d_z \vert ^1S_0\rangle$ its electric dipole amplitude
 and $\delta$ the laser frequency detuning. 
 Actually, the $^{171}$Yb isotope has two hyperfine (hf) components, $F=\frac{1}{2}$ and $F=\frac{3}{2}$ with relative strength $\frac{1}{3}$ and $\frac{2}{3}$ and hyperfine splitting $W_{hf}/\hbar = 5.939$~GHz \cite{Wij94}. It is  important to adjust the laser frequency midway  between the two hf lines, ({\it i.e.} opposite detunings, $\hbar \delta= \pm W_{hf}/2$, for the two transitions). This particular choice has the great advantage of canceling out the imaginary part of the  effective magnetic field associated with spontaneous emission (see Appendix). Only  the number of   {\it dressed} atoms, irrespective    
   of their nuclear spin orientation,  is subjected to  a decrease.
With this choice the contributions to $ \vB_{opt}$ coming from each  transition happen to be equal.
  
 It is convenient to write   $  \vB_{opt}$ in a way which makes its coupling to  the  $^{171}$Yb  ground state exactly similar to the  real magnetic field coupling.  Since it originates from an ac Stark shift, it has, however, no reason to  have a vanishing divergence.  From the calculation presented in the Appendix, Eqs. (\ref{Hopt}, \ref{opla}), we obtain the expression,  
 \bea
\vB_{opt} &=& - \frac{ 8 \,d^2}{3\, \gamma_I \,\hbar W_{hf}} \, i \,  \vE ^{*}(\vr )   \wedge  \vE ( \vr) \nonumber \\
&&\hspace{-8mm}= B_0\; \xi \,\rho\,  \Big( \sin^2(k\,x) \hat x  +x \rightarrow y 
 -\sin (k\,x) \, \sin (k\,y)  \hat z\Big), 
 \nonumber \\
 &&\hspace{-11mm} \rho  =  \frac {32 \, d ^2\, \mathcal{E}^2 }  {3\, ÊW_{hf} \,\hbar \,\gamma_I B_0 }
          =\frac{8 \; \hbar \Omega^2}{Ê3 \,W_{hf} \,\gamma_I B_0}. 
 \label{Bopt}
 \eea 
Then, the dressed atom Hamiltonian is  expressed as:
\bea
\mathcal{H}  &=&\frac{\vp ^2}{ 2\, M}  - 
\gamma_I  (  \vB_0 + \vB_{opt}(\vr))\cdot \vI +\mathcal{V}(\vr ),   \nonumber \\
\mathcal{V}(\vr ) &=&-\; \hbar \gamma_I B_0 \;\rho \;  ( \sin^2(k\,x)+\sin^2(k\,y) )/4,
  \eea
  where the scalar potential  $\mathcal{V}(\vr) $ is associated  
 with the scalar terms $\propto  \vE ^{*}\cdot \vE  $  (see 
 Eq. (\ref{Hopt})). 
   \begin{figure}
    \includegraphics[ width=8cm]{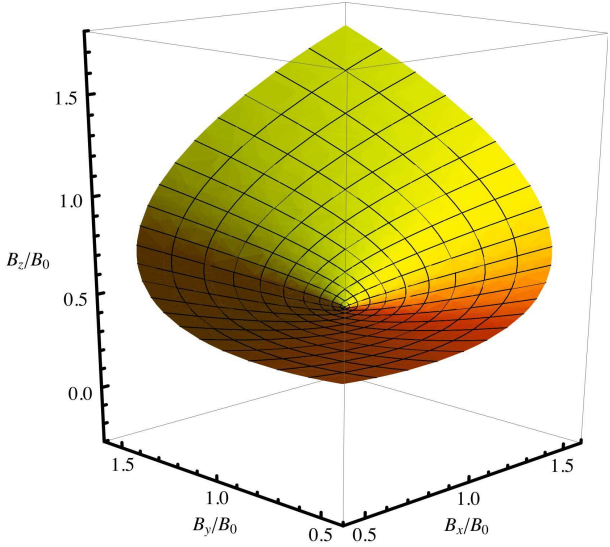}
\caption{Illustration of  the geometry embedded in the effective field $ \vB(\vr)$ by a perspective view 
 representing the 2D surface $\mathcal{S} $,  immersed in a 3D Euclidian space. $\mathcal{S} $ is defined  by the parametrization:
 $ x_1= B_x (x,y) /B_0 \, , y_1= B_y (x,y) /B_0  \,,\, z_1= B_z (x,y) /B_0$.  Performing the simple change of coordinates:
  $x_2= x_1-1/2 =\sin^2(k\,x) ;y_2 = y_1-1/2  =\sin^2(k\,y); z_2= z_1-\sqrt{1/2}= - \sin(k \,x)    \sin(k \,y) $,
  one finds   that  $\mathcal{S} $ is the intersection  of a cone of   equation: $z_2^2=x_2 \, y_2$  with the cubic 
  volume defined by $ 0 <  x_2,~ y_2 < 1$,  together with the condition $ -1 < z_2  < 1.$
 The ``rotated'' frame method is associated with the mapping of $\mathcal{S} $  onto a line segment along the $z_1$ axis. The  cone tip 
 can be  associated with the saddle point appearing in Fig. 1 between two bumps of the geometric field $\mathbf B_3$.  }
 \label{Fig2}
\end{figure}
 \subsection{Explicit  evaluation of the phases associated with the magnetic field geometry}
  In the following, the AB phase will be considered as a diagnosis.  It serves to test for the presence of a vector potential $\mathbf{A}_3(\vr)$ able to generate the non-local quantum physical effects associated with the  magnetic field $\vB_3(\vr)$.
 
  We have all we need to construct the non-Abelian gauge fields given by Eqs. \ref{objetA}. The  angles $ \theta(\vr) $  and $ \phi(\vr) $ appearing
 in the relevant unitary transformation:
 $  \mathcal{U}(\vr)  = \exp ( - \frac{i}{2} \sigma_z\,\phi(\vr )).
   \exp (-\frac{ i}{2} \, \sigma_y\,\theta(\vr ))$
  are given by the orientation of the total magnetic field, 
  $\vB(\vr) = \vB_0(\vr)+\vB_{opt}( \vr) $, 
  \bea
   \tan\phi(\vr ) &=&\frac{ B_y (\vr)}{ B_x(\vr) } \, ,\, \; \; \; \;
   \cos\theta(\vr ) =\frac{B_z (\vr)} {\vert B(\vr)\vert }.\nonumber 
\eea
  The graphs of Fig. \ref{Fig1} and \ref{Fig2}  have been obtained for  
   the typical case $  u = \pi/4 , \;\xi =\rho=1$. 
   The geometric magnetic field $ \mathbf{ B}_3 (\vr) $ can, then,  be obtained
  directly from the compact   geometric formula (\ref{B3}). 
  The  algebraic structure of $ \phi(\vr ) $ and $\theta( \vr )$
  suggests that  $ \mathbf{ B}_3 (\vr) $   can be written 
  under the   symmetric form,   
\bea
\mathbf{ B}_3 (\vr)&=& - \hat{z}\,
 k^2 \cos (k\,x) \cos (k\,y)\,\mathcal{F} \big (\sin(k x),\sin(k y) \big ), \nonumber   \\
\mathcal{F}(a,b)  &=& (a^2+2\sqrt{2} a\,b+b^2) G(a,b) ^{ -\frac{3}{2}},
\nonumber\\
G(a,b) &=& 1+ a^2 -\sqrt{2} a\,b+b^2 +
 a^4+b^4+ a^2 \,b^2 . 
 \label{Bgeom}
 \eea
   By looking at Fig.1,  it  is clearly  of interest  to compute the  phase $\Phi_{geom}(1/2) $  
     as     the   flux of $ m \,\mathbf{ B}_3 (\vr) $ through the plaquette:
 $ \pi/2 \leq   k\,x \leq 3\pi/2  $, $-\pi/2  \leq  k \, y \leq  \pi/2 $.  Using the above 
explicit  expression of $ \mathbf{ B}_3 (\vr) $, and  performing a 
  precise numerical quadrature, one obtains 
 $\Phi_{geom}(1/2)= 0.499284$.
 This pure number is a characteristic of the geometry associated with  $\vB(\vr) $ 
 for the  values of the geometric parameters $  u = \pi/4 , \;\xi =\rho=1$. 
 
 It is of interest to construct, from the  geometric field $\vB_3(\vr)$,  an electromagnetic field   which would induce the same physical effect ({\it e.g.} the same Aharonov-Bohm phase for a particle with elementary electric charge $e$). 
  This can be achieved by comparing the covariant derivative of Eq.(\ref{covD}), 
   in the limit ${\vA}_1(\vr)={\vA}_2(\vr) = 0$ (namely
$ \frac{ \partial}{\partial \vr } - i \, m \,{\vA_3}(\vr) $, $m \hbar$ being  the eigenvalue $\pm\frac{1}{2}$ of $I_z $),  with the corresponding   expression in the electromagnetic case:
   $\frac{ \partial}{\partial \vr } - i  \frac{e}{ \hbar } \,\vA_{em}(\vr)$.
  A simple inspection leads to:
      $ \vA_{em}(\vr)=m \frac{\,\hbar  } {   e}{\vA_3}(\vr)   $. 
This implies  
$ \vB_{em}(\vr)= \pm\frac{1}{2}  (\frac{ 2 \pi } {\lambda ({\rm Yb})})^2 \frac{\,\hbar  } {   e}(\mathbf{ B}_3(\vr) k^{-2}) $, 
where $ \lambda({\rm Yb})= 0.5558\times 10^{-6} m $ is the wavelength of the
 $ ^1S_ 0 \rightarrow  \,^3P_1\, ^ {171}$Yb  transition.
 Using the values of $e$ and $\hbar $,  one  gets
  the electromagnetic equivalent in $MKSA$ units of the geometric field $\vB_3(\vr)$, 
 $\vB_{em}(\vr)=\pm 0.0420287 \times 10^4$~T$ \times (\mathbf{ B}_3(\vr)\, k^{-2}$).
  
 It is important to note that in the expression of $B_{3z}$, Eq. (\ref{Bgeom}), $ \mathcal{F} (\sin(k x) \sin(k y) )  $ is a positive definite  function  of $ x ,y$. This suggests  the possibility to  confine  the Yb atoms in  2D domains 
   where $\mathbf{B}_3 (\vr)\cdot  \hat{z} \geq 0 $, by  adding to
    the  $ \mathcal{H} $  Hamiltonian an extra scalar potential,
 \be
  \mathcal{W}(\vr) = \hbar \gamma_I \; B_0      \,\mu  \,\cos (k\,x) \cos (k\,y),
  \ee
 where $\mu $ is a dimensionless positive  parameter to be adjusted.
 Let us write, in the ``rotated'' frame,   the confining part of the
 ${\widehat{\mathcal{H}}}_0 $ Hamiltonian associated with the dressed  $ (4f^{14}~ 6s^2)^1S_0,\; I=\frac{1}{2}$    ground state of $^{171}$Yb,  
\be
 {\widehat{\mathcal{H}}}_c =\frac{\vp ^2}{ 2\, M} +
      \Big(  \mathcal{V}(\vr)  +  \mathcal{W}(\vr) - 
  \gamma_I \; \vert \vB(\vr) \vert\, I_z  \Big) .
   \ee
We have omitted the gauge vector fields  $\mathbf{A}_i(\vr)$ contributions, which play a limited role in the confinement.  In  the rhs  of Fig. \ref{Fig1}, we have plotted  in red,
 using $ \gamma_I B  \hbar $ units, the total
  potential acting on the  hyperfine sub-state $ I_z=  \hbar/2 $, 
\be   
       V_c(\vr ) =\mathcal{V}(\vr)  +  \mathcal{W}(\vr) - 
\gamma_I \, \hbar/2 \,\vert \vB(\vr) \vert .
\ee
 We have  found that the choice  $ \mu= 2 $ leads  to the wanted  effect. On  the same graph, we have displayed in blue  twice the dimensionless quantity 
  $\mathbf{ B}_3 (\vr) \, k^{-2}$.  It is clear  that,
 for an appropriate temperature,  cold $^{171}$Yb atoms can be  confined in connected domains of the 2D-space  where the effective magnetic field satisfies 
  $\mathbf{ B}_3(\vr) \cdot \hat z  \geq  0  $.
   For this purpose, the experiment should make use of two optical lattices having the same periodicity, but each one fulfilling a different role. The green lattice (555.8 nm) provides the topology of the effective field which  generates  the geometric non-Abelian field and the associated AB phase. The confinement role  is attributed to the second lattice,  spin-decoupled, hence not affected by the ``rotated'' frame transformation. This provides the possibility to adjust the atom position according to their temperature. 
 \begin{figure}
    \includegraphics[ width=8cm]{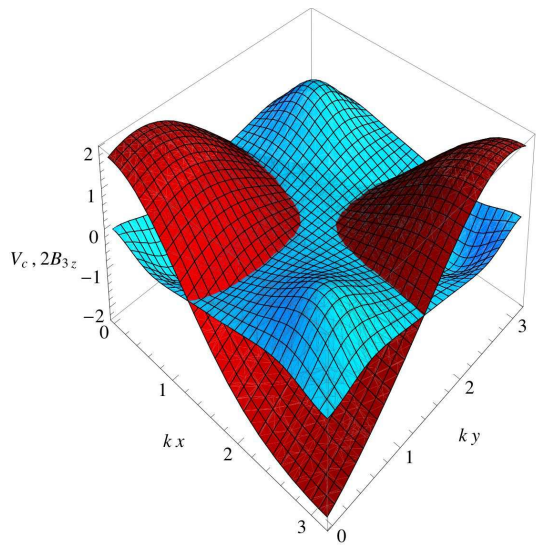}
\caption{The scalar potential $V_c(\vr)$ (dark red) and the geometric field $B_{3z}$ (light blue), within one quarter of the elementary cell relative to the potential $ \mathcal{W}(\vr)$.
The geometric phase relative to this plaquette cancels, but when $\mathcal{W}(\vr)$ is increased adiabatically,  the trajectory of the wave packet is clearly modified: the cold atom is attracted towards  the $B_{3z}>0$ regions and gets submitted to a Laplace-type force. The quantum
circuit no longer satisfies the  conditions required for the observation  the AB effect. Therefore the cancellation of the purely geometric phase no longer implies the vanishing of the magnetic flux through a physical circuit, hence cannot be considered as a negative test of the $B_{3z}$ field relevance.}
\label{Fig3}
   \end{figure}
     
One way to generate the scalar potential $\mathcal{W}(\vr)$ is to rely upon the  scalar ac Stark effect with a large detuning. This may involve  two stationary waves linearly polarized along $\hat z$,
  having wave vectors  $ {\vk}_1= k  (\hat x + \hat y)/2 $
  and ${\vk}_2= k  (\hat x - \hat y )/2$  with
  $\vert {\vk}_1\vert=\vert {\vk}_2\vert = k/\sqrt{2} $, {\it i.e.  $\lambda = 786.0$} nm.
   We make the further assumption that there is  no phase coherence between the two stationary waves, in other words, they do not interfere. With an  appropriate   phase and intensity matching  of each stationary wave,  the light induced potential, up to a constant factor, is given  by:
  $\cos (\ k\,(x+y)/2 )+   \cos (\ k\,(x-y)/2 )=  2 \, \cos ( k\,x) \cos (k \,y) $.  It has clearly the right space dependence for the wanted potential  and  the wanted sign is provided by the negative red detuning. The  light intensity can be used to adjust the potential depth without affecting the stability of the dressed atoms, thanks to the large magnitude of  the detuning $\simeq 10^{8} \, \Gamma_P$.
  
   An important property of the flux of $\vB_3$ is its cancellation when it is taken through a full elementary cell. This is an obvious consequence of the periodicity of the gauge vector field $\vA_3(\vr)$ generated by the periodic optical field $\vB_{opt}(\vr)$. 
    From a look  at Fig.1, it is clear that the symmetry  properties of $\mathbf\vB_3$  imply that its flux $ \Phi_{geom} $ through  the plaquette  now associated with  the {\it  half-elementary cell}, defined by  $ \pi/2 \leq k x  \leq 5\pi /2 \, , \,-\pi/2 \leq k y \leq  \pi /2 $, vanishes. 
  The vanishing of   $ \Phi_{geom}$ - a  {\it purely geometrical result}  -  
  is not affected by the scalar potential $\mathcal{W}(\vr)$ which has no effect upon  the  ${\vA}_3  ( \vr ) $
  gauge field. In the limit where  the kinetic term dominates $   V_c(\vr ) $  in the Hamiltonian $ {\widehat{\mathcal{H}}}_c$, it is possible to build wave packets which propagate along the plaquette  perimeter. The AB  phase,  $\Phi_{AB}$,   
  around the half-elementary cell will vanish since it coincides  with $ \Phi_{geom} $. 
     However, if we assume that  $\mathcal{W}(\vr)$ is increased adiabatically, the
wave packets are pushed towards the regions where $B_{3z}>0$. Depending on their kinetic energy, they
may even be confined there.   In these conditions, the phase accumulated
along the deformed circuit will differ from $\Phi_{geom}$ for two
reasons: first,  the $ B_{3z} $ flux  is no
  longer guaranteed  to vanish;   second, the wave packets propagate  in regions where $ \vert B_{3z} \vert  > 0 $ which makes the situation incompatible with physical requirements for measuring the AB effect, see Fig. \ref{Fig3}.  Therefore the vanishing of the $B_{3z}$ flux through this particular purely geometric loop cannot be considered as an evidence for the irrelevance of the  $\vB_3$ field in presence of a confining potential.     
       \begin{figure*}  
 \includegraphics[height=9.0 cm]{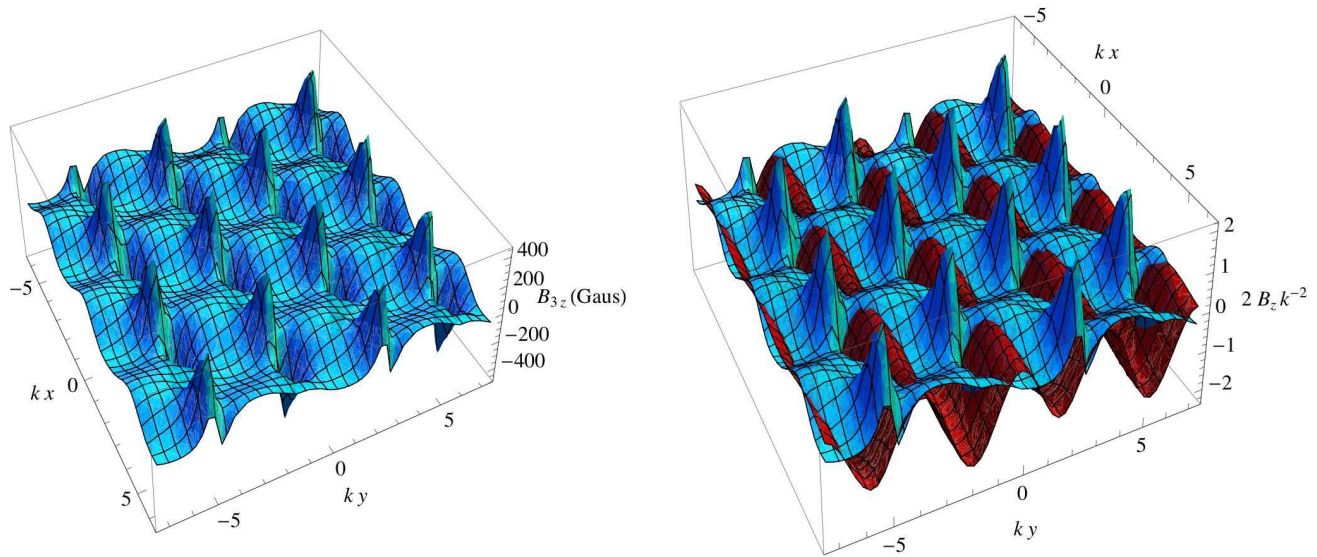}
\caption{The geometric field  $\vB_3(\vr) $ (in light blue) and the scalar confinement potential (in dark red)  for  the triangular optical lattice 
within the rotated frame. {\it Left} : The holes and the bumps
of   $\vB_3(\vr) $  form an hexagonal  lattice. They can be organized in rows alined along the $x$-axis.  {\it Right} : This can be used to introduce in $\widehat{\mathcal{H}} $  an extra $y$-dependent $\mathcal{W}(\vr)$ potential leading to the total scalar potential  $ V_c(\vr ) $  acting on the state $ I_z = \hbar /2 $ (in 
  $\gamma_I \hbar \,B^{tri}_0$ units). For sake of visibility $ V_c(\vr )$ is shifted up by 1 unit. }
 \label{Fig4}
 \end{figure*}
   \subsection{The instability of the dressed atom versus  the transverse field corrections}
 By contrast to what happens for $\vB_{opt}$, the  imaginary part of the scalar potential  $\mathcal{V}(\vr) $  does not cancel out. This leads to  an instability of the ``dressed" atom associated with  the spontaneous decay of the mixed excited state \cite{bou2,cct}.
 However,  the long lifetime $\Gamma_P^{-1}= 850$ ns of the $^3P_1$ atomic state \cite{Bowers} contributes to slow down the decay of the dressed  atom  governed   by the rate $\Gamma^{\prime}= (4 \hbar^2 \Omega^2/W_{hf}^2) \Gamma_P$ for the chosen detuning (see Appendix). There is a clear  relation between the magnitudes of $B_{opt}$ and $\Gamma^{\prime}$ since they are both proportional to $\Omega^2$, hence to the laser intensity $I_l$.
 Following the current use, the laser intensity is defined in terms of the saturated intensity $I_s$ at which $\Omega^2 = \Gamma^2/2 $. For the Yb transition considered, $I_s$ amounts to 0.14 mW/cm$^2$ \cite{Yabu99} leading to a decay  rate $\Gamma^{\prime}= 1$ s$^{-1}$, obtained for a laser intensity $I_l$= 73 mW/cm$^2$ which gives rise to $B_{opt}= 4.824$ G. Obviously, the laser intensity will have to satisfy a compromise if one wants to keep $\Gamma^{\prime} \propto I_l$ small as well as the parameter $\eta\propto I_l^{-2}$ which scales the importance of the transverse gauge field  corrections. Below we show that a realistic compromise does exist, allowing one to anticipate a magnitude of corrections at the percent level.

 To get a more precise  evaluation of the  transverse gauge field corrections than
 the one given in equation (\ref{eta}), we have calculated numerically the average of the two quantities  $ \mathcal{ N} =\sum_{i=1,2}(\frac{\partial}{  \partial  \vr  } \cdot\vA_{i}( \vr))^2 $ and    $\vert B(\vr)\vert $, using their exact expressions over the relevant plaquette.
In the rhs of Eq. (\ref{H1sqav}), the second term, once averaged, is found to be ten times smaller than the first one, whose average leads to $ \mathcal{N}_{av} =1.53445 ~k^4 $,
   while ${\vert B(\vr)\vert}_{av}= 1.684~B_0$. From their ratio and the value of $\eta_0= \frac{1}{2} \Big ( \frac { (\hbar \vk)^2 /(2 \,M ) }{\hbar \gamma_S  B_0} \Big ) ^2 =12.64 $  for $^{171}$Yb and for $B_0=1$~gauss. We deduce $\eta_{av}=  \eta_0  \, \mathcal{N}_{av}/{\vert B(\vr)\vert}_{av}= 6.828~B_0^{-2} $ (for $B_0$ expressed in gauss). This gives an improved   estimate  of  $\eta_{n,m}$ (Eq. (\ref{eta})) in this well specified geometry,  assuming that the Bloch wave functions moduli are slowly varying over one elementary cell.  One can reduce  $\eta_{av}$  to the one percent level, by  choosing  $B_0= 24$~G. For the total  laser intensity this leads to $I_l = 365$ mW/cm$^2$, shared among the four beams,  with ${\Gamma^{\prime}}^{-1}= 0.2$~s for  the  ``dressed" ground state lifetime.  
  \subsection{ Extension to a triangular  optical lattice involving running waves}
     To end this section, we would like to present  an extension of the above analysis
   to the case  a triangular optical lattice,  generated  by a set of three interfering running waves, invariant  by rotation of  $  2 \pi/3 $-multiple angles. The triangular geometry has been observed for the first time by Grynberg {\it et al.} \cite{Gry93} and the properties of cold atoms trapped in such a 2D lattice  have already been considered in \cite{Dud04} but with a spatial configuration different from the one adopted here. In contradistinction to what we do, the authors  have not taken into account the two hyperfine transitions.  
   
We consider three circularly polarized, running waves, which propagate in the $x,\,y$ plane. It should  be noted that  the previous square  lattice  built  from stationary waves cannot be identified with  a running wave  lattice invariant by rotation 
   of $ \pi/2$-multiple angles,   since the   normal  reflexion of an optical wave cannot be described by  a  rotation of $ \pi $ of the wave electric field.    
   Our optical triangular  lattice  is described by the following optical electric field,  
   \bea   \vE(\vr) &= &\mathcal {E} \sum_{n=0}^{n=2} \,\ve_n(\xi ) \exp (i \vk_n  -i \zeta (n) ) ,\nonumber  \\
   \vk_n & =& R(\hat{z} , 2 n \pi/3). \hat{x} ,   \nonumber  \\
   \ve_n(\xi )&=&   R(\hat{z} , 2 n \pi/3). ( \hat{y} + i\xi \hat{z})/\sqrt{2} . 
   \eea 
            It turns out that  the choice   $ \zeta (n)= 2 n \pi /3 $  is  slightly more favorable  than  $\zeta (n)=0 $. 
      To simplify the writing, it  is  convenient to  introduce  the dimensionless vector field: 
      $ \vb(\vr )=   -  \frac{ i }{ 2 \, \mathcal {E}^{2} } \vE ^{*}(\vr )   \wedge  \vE ( \vr)     $. 
          Performing a  straightforward  computation using the above expression  of $\vE(\vr) $,  one arrives  at the rather compact expressions for   the  $ \vb(\vr )$ field  components:
 \bea
          && \hspace{-5mm}     b_x= \frac{\xi}{2} \Big(-\cos \big(\sqrt{3} y - 2\zeta\big)+\cos \big(\frac{3 x}{2}\big) \cos
   \big( \frac{\sqrt{3} y}{2}- \zeta\big)\Big) ,\nonumber \\
     && \hspace{-5mm}  b_y =  \frac{\xi}{2} \sqrt{3} \sin \big(\frac{3 x}{2}\big) \sin \big(\frac{\sqrt{3} y}{2}-\zeta\big) ,\nonumber \\
    && \hspace{-5mm}   b_z =\frac{\sqrt{3}}{2} \sin \big(\frac{\sqrt{3} y}{2}-\zeta\big)
       \Big(\cos \big(\frac{3  x}{2}\big)-   \cos \big( \frac{\sqrt{3} y}{2}-\zeta\big)\Big).   \nonumber\\
\eea
   By contrast to the previous case, $b_z$ is independent of  the beam helicity.
  Using  equation (\ref{opla})  the optical effective magnetic field is  given by:
   \be
   \vB_{opt}^{tri} =  \frac{ 16 (\mathcal {E}\,d)^2} {3 \gamma_I \,\hbar W_{hf}}  \vb .
   \ee    
  As before, in  order to avoid singularities  in the geometrical gauge  fields,  we  have 
  added to $\vb(\vr )$  a  constant field: $ \vb_0 = \sqrt{2}  \, \hat{x} + \frac{1}{2}\hat{y} $, leading to a  constant magnetic field 
   of magnitude $ B_0^{tri}=  \frac{ 8 (\mathcal {E}\,d)^2} { \gamma_I \,\hbar W_{hf}}   $. Following the procedure used in Sec. V. B for the square lattice, we have calculated the geometric field $\vB_3(\vr)$ satisfying the formula (\ref{B3}). 
  On Fig. \ref{Fig4}, the {\it Left} graphic gives a perspective 3D plot  of $\vB_{em}(\vr) $, the electromagnetic field equivalent  of the geometric field $\vB_3(\vr)=
 \hat{z} \cdot\frac{ \partial}{\partial \vr }\wedge {\vA}_3(\vr)$,  as a function of   $ k \,x $  and $ k \,y $.   
   The geometric   phase $\Phi_{geom}(1/2) $,  
     given   by   the   flux of $ m \,\mathbf{ B}_3 (\vr) $ through the plaquette   
  $0\leq   k\,x \leq   4 \pi/3 $ cancels, but through the half-plaquette $0\leq   k\,x \leq   2 \pi/3 $ it amounts to  $ \Phi_{geom}(1/2)= 0.384137 $. Larger values are obtained by deforming the plaquette, {\it e.g.} for
 $ -0.9 \leq   k\,x \leq  1.92   $, $-0.2  \leq  k \, y \leq 2.8 $, it becomes $ \Phi_{geom}(1/2)= 0.475858 $.
 We have calculated the corrections due to the transverse gauge fields by following the procedure already described for the square lattice. The result $\eta_{av}$ = 6.86 ($B^{tri}_0)^{-2}$ turns out to be close to the one obtained in this previous case (6.83~$B_0^{-2})$. 

 The geometric magnetic field $ B_{3z}$   exhibits regularly spaced pairs of positive maxima and negative minima   forming  hexagonal lattices. It appears that the bumps and  the   holes can be  organized in rows parallel to the x axis. This offers the possibility to bar  cold $^{171}$Yb  atoms from  visiting the 
  connected domains of the 2D-space  where $ B_{3z}<0 $, by adding to   
     the  $ \mathcal{H} $  Hamiltonian the following  $y$-dependent scalar potential, 
 \be
  \mathcal{W}(\vr) = \nu \hbar \gamma_I \; B_0^{tri}   \cos \big(\sqrt{3} k \,(y -y_0)   \big).
  \ee
   The confining potential plotted in red  on the right graphic of Fig. \ref{Fig4} corresponds
    to the parameter choice  $ \nu =1.$ and $y_0=0$. 
  
 Challenging experiments are currently performed by overlaying two commensurate triangular optical lattices generated by light  at the wavelengths of 532nm and 1064nm \cite{Stamp}. Different 2D lattice geometries have been observed by tuning the relative positions of the two lattices. In this realization  the beams are linearly polarized, so that the periodic optical magnetic field has a fixed direction, normal to the lattice, leading to a null gauge magnetic field $\mathbf{B}_3(\vr)$. Our work paves the way to predicting and designing space-dependent optical fields which will endow eventually  trapped fermions with orbital magnetism, the magnetic spin quantum number of the trapped atoms playing the role of their electric charge.  
\section{Summary and Perspectives}
The purpose of this paper is the study of the geometric  phases emerging
from the non-trivial topology embedded in space-dependent vector fields
$ \vB(\vr) $, coupled to the spin of a neutral particle. Instead of working directly on
the geometry of the  spin wave functions, we  have relied, in analogy with our previous
work \cite{bou1}, on  a ``rotated frame"  approach, where a space-dependent unitary
transformation $ \mathcal{U}(\vr) $ is applied to line up
the local  $ \vB(\vr) $  field along the spin quantization axis:
  $ \mathcal{U}^{\dag}\, \vS \cdot \vB \,\mathcal{U} = \vert \vB\vert\, S_z $.
Introducing the standard Euler angles $ \theta(\vr) ,\phi( \vr) $, giving the direction
of $\vB $,   the wanted transformation reads: $\mathcal{U}(\vr)  = \exp (\frac{- i }{\hbar}\, S_z\,\phi).  \exp (\frac{- i}{\hbar} \, S_y\,\theta) $. In the ``rotated"  Hamiltonian
    $\widehat{\mathcal{H}}=  \mathcal{U}^{\dag}(\vr)  \,  {\mathcal{H}}\,\mathcal{U}(\vr)$, appears the non-Abelian gauge  field        $\mathcal{A}(\vr)= i \hbar \,\mathcal {U}^{\dag} \cdot \frac{ \partial}{\partial \vr }\, \mathcal{U}  $  which encapsulates  the geometry of $ \vB(\vr)$:
$\widehat{\mathcal{H}}={ \widehat{\vp} }^2 /(2 \,M)+V(\vr )+\vert \vB\vert\, S_z$, the `` rotated"  momentum $\hat \vp$ being  given by the significant  expression,
 $\widehat{\vp} =  -i \hbar\frac{ \partial}{\partial \vr }-  \mathcal{A}(\vr) $.
 Relying  only on the $SU(2)$  Lie algebra: $[ S^i,S ^j]=i\,\hbar \, {\epsilon}_{i\,j \,k} \, S^k $,    $ \mathcal{A}(\vr) $  can be written as  a linear combination  of
$ S^1=S_x, \; S^2=S_y, \; S^3=S_z $,  namely $\mathcal{A}(\vr) = \sum_{k=1}^3 S^k \,\vA_k(\vr) $.
   The   three vector  fields  ${\vA}_i(\vr)$  involve the  gradients of $ \theta(\vr) ,
   \phi( \vr) $ (Eq. (\ref{objetA})), in particular the ``longitudinal" vector field is $  {\vA}_3(\vr)= \cos\theta(\vr)\,\frac{ \partial}{\partial \vr }\phi(\vr) $.

   Performing the non-Abelian gauge transformation $ \widehat {\Psi} \rightarrow
 \exp (\frac{i}{\hbar} \Lambda_{k}(\vr)\, S^k  ) \widehat {\Psi}$
  leaves the Hamiltonian $\widehat{\mathcal{H}}$ invariant,
      if   $\mathcal{A}(\vr) $    transforms like
   a  standard SU(2) Yang-Mills field. However,  $\mathcal{A}(\vr) $
  differs from  a Yang-Mills field in an  essential way: it has a vanishing  non-Abelian
``curvature".  This  means that, unlike $SU(2)$  Yang-Mills fields,
 $\mathcal{A}(\vr) $ is a  purely  geometric object, not a dynamical
   self-interacting  non-Abelian  gauge field.

 Pursuing  the analogy with Berry's phase, we have  constructed a  Born-Oppenheimer-like perturbation  scheme: $\widehat{\mathcal{H}}= \widehat{\mathcal{H}}_0 +\widehat{\mathcal{H}}_1$, which takes the place of the adiabatic approximation with its corrections. The Hamiltonian   $\widehat{\mathcal{H}}_0 $  contains  only the ``longitudinal"  gauge vector field
  $  {\vA}_3(\vr) $ which is assumed to play   the dominant role
 while the ``transverse"  fields  $  {\vA}_{1,2}(\vr) $,  incorporated in $ {\mathcal{H}}_1 $,  generate well defined BO corrections, appearing only to second order. The    spin dependence   of  $\widehat{\mathcal{H}}_0$ reads  $\big(\frac{ i\hbar}{ 2\, M}
 \{ \frac{ \partial}{   \partial  \vr }, {{\vA}}_{3} (\vr) \}
   +\gamma_S \vert \vB(\vr)\vert\, \big)\, S_z .$
(At this stage of our work, we   limit  ourselves to  a $ S=1/2$  particle in a 2D space.)
The role of Berry's phase is taken up  by
 the Aharonov-Bohm  phase $\Phi_{A B}(m) $  given  by  a loop  integral of   the  vector   gauge field    $  {\vA}_3(\vr) $  or by  the 2D  flux integral     of  the  associated ``geometric   magnetic" field:
 $   \mathbf{ B}_3 (\vr) =      \sin(\theta(\vr) ) \Big ( \frac{ \partial}{\partial x }\phi(\vr) \, \frac{ \partial}{\partial y }\theta(\vr) -\frac{\partial}{\partial y }\phi(\vr) \, \frac{ \partial}{\partial x  }\theta(\vr) \Big ) \,  \hat z$,  
the eigenvalue $m$  of $ S_z/\hbar $ standing for the ``effective" electric charge.
Going back to the laboratory frame, the explicit evaluation of $\Phi_{A B}^{lab}(m) $  is performed in sec. III within the BO approximation, where $\widehat {\mathcal H} \approx \widehat {\mathcal H}_0 $. In a typical experiment, both interfering paths remain in regions where $\mathbf{ B}_3 (\vr)=0$.  
 Then, the Feynman phase factor can be factored out from the wave function $\widehat \Psi(\vr) = \widehat{ \psi}_{i} (\vr,t) \otimes \widehat{\chi}_i(\vr)$ and the evolution of the  spatial part $\psi_i(\vr)$  ruled out by the Hamiltonian  $\widehat {\mathcal H}_0 (\vA_3 =0)$, is just identical to that of $\widehat \psi_i(\vr)$. Thus, the change of phase is only governed by the spin transformation $\chi_i(\vr_i) = \mathcal{U}(\vr_i) \hat \chi_i $. At the interference point, $\vr_1= \vr_2$, unitarity of  $\mathcal{U}(\vr)$ simply implies that $\Phi_{A B}^{lab}(m) = \Phi_{A B}(m) $.
In addition, we give an explicit method to estimate the transverse field corrections to
$\Phi_{A B}(m) $. They  are given in terms of a set of parameters $ \eta_{n,m} $ which can
be viewed as the squared amplitudes of the first  order ${\mathcal{H}}_1 $  corrections to the eigenfunctions of $  \widehat{\mathcal{H}}_0 $. If
  $ \vB(\vr) $  is a periodic field  with a single   wave number  $\vk$, we have  derived for  $ \eta_{n,m} $  the following  estimate, 
    $ \vert\eta_{n,m} \vert \lesssim   1/8 \Big (  (\hbar \vk)^2 /
    ( \hbar \gamma_S \,M {\vert \vB(\vr)\vert}_{min} \Big ) ^2 $.

In order to illustrate the above conceptual developments on a concrete physical case, we have applied our formalism to cold $^{171}$Yb atoms dressed by the beams of an optical lattice. The dressing optical field involves two stationary waves directed along $\hat x$ and $\hat y$, each beam carrying the angular momentum $\xi \hbar  $ per photon. The light frequency is tuned midway between the two transitions connecting the $(6s^2)^3S_0$, $F=I=\frac {1}{2}$ ground state to the hyperfine sublevels $F=\frac {1}{2}$ and $\frac {3}{2}$ of  the $(6s6p) ^3P_1$ excited state.  This choice guarantees that the dressed atom instability does not affect the effective dc magnetic field  acting on the atom (see \cite{bou2} and Appendix). A real static magnetic field of comparable magnitude is added to avoid possible singularities of the optical gauge field. This geometry is simple enough to lead to a tractable analytical expression of the rescaled geometric field, $k^{-2} \vB_3(\vr)$, which depends {\it only} on the geometric properties of the system Hamiltonian, namely relative directions of propagation axes and polarizations of the beams but is {\it independent of the light intensity}. 
 This latter only determines  the magnitude  of the corrections associated with the transverse gauge fields via $\mathbf \vB(\vr)$. The explicit expression of $\vB_3(\vr)$ can be
casted under the following form: $ \vB_3(\vr) = -\cos(k x) \,\cos(k x)\mathcal{F} \big (\sin(k x),\sin(k y) \big ) k^2\,\hat{z}$,  where $\mathcal{F} $ is a positive definite function
of  $ x$ and $y $. This offers the possibility, within the rotated frame,
 to confine the  $^{171}$Yb atoms in 2 D domains where $\hat{z}.\vB_3(\vr)$
  is positive by introducing an extra scalar
potential $\mathcal{W}(\vr) = \hbar \gamma_I \,B_0 \,\mu  \,\cos (k\,x) \cos (k\,y)$,
as illustrated on  Fig. \ref{Fig1}.

 We have added  at the end of this section a short presentation of the geometric magnetic field  $ \vB_3(\vr) $
  emerging from the geometry  of the optical field  $ \vB_{opt} $   generated  by a triangular  optical  lattice. We consider a lattice 
   resulting  from  the interference of  three circularly polarized running waves,  
  invariant  by rotation of     $  2 \pi/3 $ multiple angles.
 The associated  $ \vB_3(\vr) $ field exhibits a 2D hexagonal geometrical pattern involving bump and hole pairs. Like in
the case of the square lattice generated by stationary waves, it is possible to confine the  $^{171}$Yb atoms outside the  $ B_{3 z} $ holes,  as shown on  Fig. \ref{Fig4}. 

    This work is based on the sole geometric properties of the Hamiltonian. Our initial problem, formulated in the rotated frame, has been reduced, within a BO-like approximation, to that of  a spinless  particle, having an electric charge $ \propto m $, interacting  with periodic Coulomb  and magnetic
   vector potentials, {\it i.e }  a standard - though non trivial - problem of Solid State Theory. What remains to be done, with our concrete example, is to fully determine  the energy bands and the associated Bloch wave functions and explore directly their geometry. This program is beyond the  scope of this exploratory paper.  
   
   \section*{Acknowledgements}
   We are grateful to Jean Dalibard and Nigel Cooper for reading our paper and making stimulating comments.  

\section*{Appendix : Optimal detuning of the dressing beam}
When the light detuning is of the order of the hyperfine (hf) splitting, the Hamiltonian which acts on the optically dressed $^1S_{0, F=\frac {1}{2}} \; ^{171}$Yb ground state, receives distinct contributions from the two hf lines.
 We derive them by assuming here for simplicity, that the light wave is a running wave so that the light angular momentum  $ \propto \xi \hat z= i \vE^{*} \wedge \vE/\vE^{*} \cdot \vE $  is spatially uniform.

There are two remarks which simplify the calculation.
 
1) For the $ 1/2 \rightarrow 1/2$ hf line the electric dipole and the spin matrix elements are proportional, (Wigner-Eckart theorem).  The contribution to the Hamiltonian  reads $ H^{(1)} = \frac {\hbar \,\Omega^2/3}{\delta_{1} + i \Gamma_P /2}  (1 - 2\, \xi  I_z )$.

2)  For the $ 1/2 \rightarrow 3/2$ hf line, one should note that the sum over the excited states involving the projection operator ${\small \mathcal{ P}_{F=3/2}=\sum_{m_F} {\vert  F=\frac {3}{2}, m_{_F} \ket \bra F= \frac {3}{2}, m_{_F} \vert}}$  
 adds up to ${\small \mathcal{ P}_{F=1/2}}$ to give the unit operator. As a result for very large detunings off the two hf lines, the vector coupling contributions nearly cancel each other while the scalar ones add up to 
$\frac{\hbar \,\Omega^2}{\delta}$. Hence the contribution of the second hf line is
$ H^{(2)} = \frac {2\hbar \, \Omega^2/3}    {\delta_{2} + i \Gamma_P /2}  (1 + \xi  I_z )$.

 By contrast, for opposite detunings  $\delta_{1}= -\delta_{2}= W_{hf}/2 \hbar$ the vector coupling contributions become equal and add up together, while the scalar ones subtract. The complete Hamiltonian becomes:  
\noindent $ H_{opt}=H^{(1)}+ H^{(2)}= - \frac {2\hbar ^2\,\Omega^2}{3 W_{hf}}(1+4 \xi  I_z) - i \hbar \Gamma^{\prime}/2$, with $\vB_{opt} = \frac {8 \hbar \Omega^2}{3 W_{hf} \gamma_I } \xi \, \hat z$, the optical field. The imaginary part ${\small \Gamma^{\prime}/2= \frac {4 \hbar^2 \Omega^2}{\,W_{hf}^2} \; \Gamma_P/2}$ is free of $I_z$ contribution.   

 The calculation can be generalized to the case of the stationary wave given by Eq. (\ref{statwave}). 
When the two detunings are opposite, the effective Hamiltonian  is now  given by:
 \be 
   H_{opt} =     \frac{ 8 \,d^2}{3 W_{hf}}\Big(  -\frac{1}{4} \,\vE ^{*}\cdot \vE  + 
   \frac{ i}{\hbar} \,  \vI   \cdot  \vE ^{*}   \wedge  \vE  \Big).
   \label{Hopt}
   \ee 
   The  effective magnetic coupling, associated with the term $ \propto \vI $, leads to the following 
   expression for $\vB_{opt} (\vr) $: 
   \be
   \vB_{opt} (\vr) =- \frac{ 8 \,d^2}{3\, \gamma_I \,\hbar W_{hf}} \, i \,  \vE ^{*}(\vr )   \wedge  \vE ( \vr).
   \label{opla}
   \ee 
   
  The calculation of $\Gamma^{\prime}$ is also valid for stationary waves, assuming that
 $\mathcal{V(\vr) \propto \vE(\vr) ^{*}\cdot \vE(\vr)} $ can be replaced by its average over one unit cell.

  \end{document}